# Automated control strategy for setting and stabilization of photonic circuits

Jacek Gosciniak

[1]Institute of Microelectronics and Optoelectronics, Warsaw University of Technology, Koszykowa 75, 00-662 Warsaw, Poland

Corresponding author: jacek.gosciniak@pw.edu.pl

**Abstract**

In this paper, we propose the Wheatstone bridge configuration for enabling real-time and closed-loop stabilization and calibration of photonic devices integrated on chip. The measurement of the optical power propagating in a waveguide is achieved by leveraging the photo-thermal resistance variation of one of the resistors that comprise a bridge, which is either part of the waveguide or in close contact with the waveguide. The voltage generated by a monitor is applied through a processing unit to preceding actuators with a view to locking a system at the desired signal level. As all the "resistors" consisting of the Wheatstone bridge are placed on a single material platform in proximity to each other, the bridge is insensitive to temperature variations. Consequently, it functions exclusively as a monitor for the optical power propagating in a waveguide. Due to the high sensitivity of the monitor, automatic re-tuning of the Mach-Zehnder interferometer can be achieved with a recovery time defined by the material properties of the bridge "resistor". The material platform and arrangement can be implemented for both monitoring and activating the actuators, which makes the proposed system an attractive candidate for closed-loop control of optical devices. Furthermore, in contradistinction to the majority of monitors and photodetectors, which provide electrical current as an outcome of measuring an optical signal, the proposed circuit provides voltage, thereby eradicating the necessity to convert current to voltage at a later stage. This serves to firstly streamline the entire circuit and secondly contribute to a substantial diminution in the noise level in the circuit.

**Introduction**

Integrated photonics is a key technology for optical communications that is rapidly evolving for applications in sensing, metrology, signal processing, and computing [1]. Programmable photonic integrated circuits (PICs) [2, 3] are the optical analogue of field programmable gate arrays (FPGAs) [4] and can implement arbitrary filters [5, 6] and passively compute matrix operations on optical modes [7]. Like FPGAs, they can be flexibly reconfigured by software after fabrication [2, 3, 8-10]. These circuits are capable of operating on up to tens of optical modes and have been used to accelerate tasks in signal processing, combinatorial optimization, mode unscrambling, quantum simulation, and artificial intelligence [11, 12].

Programmable photonic integrated circuits (PICs) require scaling up hundreds or thousands of optical paths (modes), which involves precise fabrication of tens of thousands of optical interferometers. However, static component errors induced by process variation grow exponentially as the system expands, which limits the usability of the system. For example, a small beam splitter variation of only 2 % may degrade accuracy by nearly 50 % for feedforward circuits [**13**]. Thus, hundreds of thousands of optical paths need to be monitored through control of the actuators. It may be realized by monitors or photodetectors, which are key components of the configuring techniques, embedded immediately after or even inside each tunable element [**14**-**16**]. In consequence, the monitor signals may serve to implement optimization or self-calibration through actuators featured on the chip [**15**-**17**].

Such detectors or monitors should be placed in any part of the photonic circuit to monitor in real time an optical signal in a system [**16**]. They should operate under low optical power requirements just to sample the small amount of power, which can be convenient for calibrating or setting up the network. Up to now, the state-of-the-art monitoring devices are based on "mostly transparent" detectors [**18**, **19**] in the waveguide or can use waveguide "taps" to sample a small amount of power from the guide to conventional detectors or cameras [**20, 21**].

The transparent photodetectors usually utilize the contactless integrated photonic probes (CLIPPS) and operate based on capacitively-coupled conductance measurements in semiconductor waveguides [**18**, **19**]. Usually they require some additional doping of the waveguide core, which, however, increases the waveguide attenuation. The physical effect associated with this photodetector relies on the change of the waveguide conductance induced by the native interaction of photons with the intra-gap energy states localized at the Si-SiO$_2$ interface. It uses a capacitive access to the waveguide, thereby avoiding direct contact with the waveguide core, so no specific treatments at the waveguide surface need to be done.

In a case of monitor taps, a small fraction of the light is tapped from the waveguide using a short directional coupler and connected to a grating coupler that is used as a monitor. Usually, the directional coupler is designed to have 1 % power coupling efficiency, while an external infrared camera is used to simultaneously detect the light emitted by the grating coupler monitors of the many unit cells [**21**]. The monitor grating couplers, used for the feedback readout of the circuit, can be grouped together in a small area, which makes it possible to read all the monitors in a single frame of an IR camera.

An alternative approach relies on an implementation of an in-line Surface Plasmon Detector (SPD) that exploits the photothermal effect to monitor the optical power in a waveguide. It leverages the resistance variation of a plasmonic strip that is in contact with the waveguide and, thus, with a propagating mode. A small portion of the guided optical power is absorbed by the metal stripe and causes a local temperature rise that can be detected by probing the temperature-dependent resistance of the electrode. The responsivity for such a system was experimentally validated at 7.5 µV/mW at a wavelength of 1550 nm [**22, 23**]. With such monitoring techniques of internal powers in the mesh, simple progressive techniques allow mesh calibration by power maximization based on shining in specific vectors of amplitudes [**24-27**].

Here, we propose a control strategy to automatically adjust the optical power and the phase of photonics components to control the optical signal at each point of the photonic system with extreme accuracy and minimal insertion loss. The strategy is calibration-free, real-time, and seamlessly applicable to both MZI-based and RR-based processors. The technique is based on a feedback control loop [**22, 28**] that simultaneously adjusts the matrix coefficients of the device transfer function and compensates for process tolerances and thermal drift in real time.

The proposed control strategy for automatically configuring each photonic device relies on the Wheatstone bridge arrangement [**29, 30**] with a calibration-free feedback loop that does not require prior knowledge of the device transfer function. In this way, it may stabilize the working point of PICs against thermal and time-varying drifts, thus counteracting unwanted variations from the target.

The control algorithm to adjust the optical power at the output of each photonic device, needed to switch on/off the desired optical paths, is implemented through the Wheatstone bridge and relies on the signal voltage measurements, which are then applied to any preceding or even following actuators on the chip. This defines a feedback strategy that is easily scalable to multiple devices and independent of the average light power in the chip and the setup temperature.

**Results**

The present description relates to an automated control strategy to set and stabilize the operation of the photonic components, for example, an integrated optical gate implemented with a Mach-Zehnder interferometer (MZI). More specifically, it configures the operation at any desired working point without any prior calibration or complex algorithm for the correction of hardware non-idealities. For instance, in some cases, variations in temperature cause undesirable variations in the transmission of light through photonic components, resulting in a disorderly distribution of light inside a photonic integrated circuit. The circuitry described herein provides for high-accuracy detection of the inaccuracy and high-precision modulation of one or more characteristics of the photonics components to correct for the undesirable effects.

Besides, many other interesting phenomena are possible with a proposed arrangement, such as, for example, a stabilization of the optical signal in one of two stable states, 0 or 1, where 0 means no transmission and 1 means a maximum transmission. Other possible phenomena involve a constant oscillation of the signal between two states.

It is realized through the monitoring module that may apply the measured signal voltage to any preceding actuator, i.e., the phase shifting element or modulator, to modulate the optical signal in the monitoring module.

**Proposed concept**

Many photonic components are very sensitive to fluctuations in various parameters, for example, temperature, which leads to unpredictable results and drops in performance. As emphasized above, even a small beam splitter variation of only 2 % may degrade accuracy by nearly 50 % for feedforward photonic circuits [**13**]. Thus, precise control of those parameters is of huge interest.

Traditional thermal detectors on semiconductor materials are one of the solutions; however, those devices are characterized by a low level of precision, mostly due to the very high electrical resistivity of semiconductors [**31**, **32**]. The measurements of the performance of an optical system with such detectors involve detection of current fluctuations in the nano-ampere range. However, this signal level is below even the noise level in such systems, which causes the signals to be buried in the noise. This makes the signals impractical to use directly in electrical feedback circuits, even with the support of amplifiers that are often utilized to control a system.

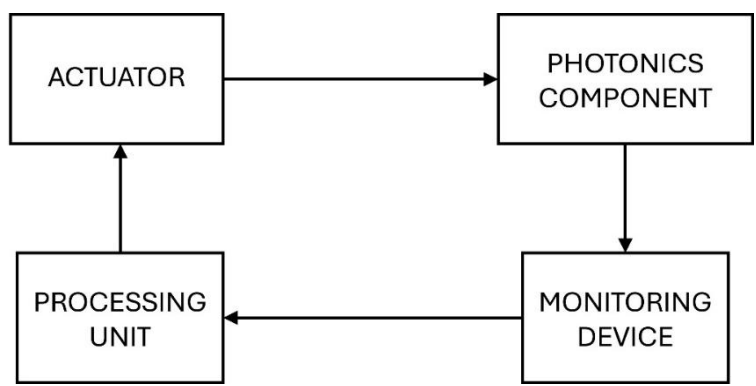

**Figure 1**. Block diagram of a system for stabilizing photonic components.

To overcome some of those limitations, the monitoring device in the form of the Wheatstone bridge configuration was proposed here. The power monitoring with the Wheatstone bridge configuration relies on the resistance measurements of the material supporting the photonic or plasmonic mode, where the absorption losses of the propagating mode cause an increase in the material temperature and, in consequence, its resistance [**29**, **30**]. As all elements of the Wheatstone bridge are arranged on the same chip, it allows reducing the influence of environmental temperature fluctuations.

**Fig. 1** illustrates a block diagram of a system for stabilizing and setting one or more aspects (e.g., temperature, power, phase) of a photonics component. An example of a photonics component, described in more detail below, is a Mach-Zehnder interferometer and ring resonator. The system includes a monitoring module configured to measure a characteristic of interest of the photonics component that may vary as a function of time. Here, a monitoring module is configured to detect a signal voltage representing power being transmitted in a waveguide. The system also includes a processing unit that performs simple operations on a signal voltage, such as subtraction of the measured voltage from a target voltage and integration of the resultant voltage to update the heater voltage of the actuator. The actuator modulates the optical signal transmitted through the photonics component to a desired value measured by the monitoring module.

In comparison with standard detectors, a proposed monitoring module that operates based on the Wheatstone bridge configuration provides information about the signal voltage $V_s$ drops across the bridge that is proportional to both the bias voltage $V_b$ and the optical power dissipated in the monitored part of the waveguide. Thus, for a constant bias voltage $V_b$, the Wheatstone bridge is capable of providing a signal, i.e., voltage, proportional to the degree of detuning of the photonics component. Therefore, the bias voltage can be used here to control the signal voltage [**29**, **30**].

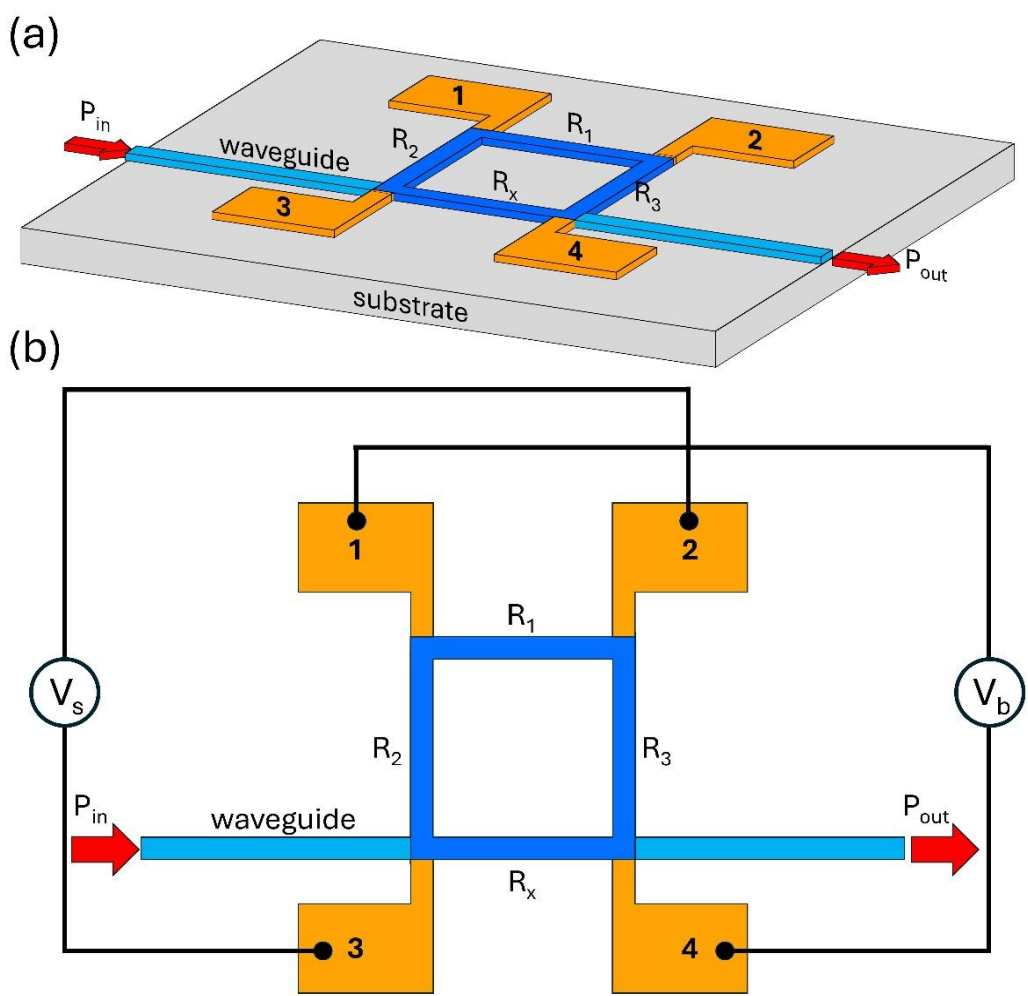

**Figure 2**. (a) Perspective view and (b) top view of the proposed photonics-based power monitor using an internal Wheatstone bridge configuration.

**Fig. 2** illustrates a Wheatstone bridge composed of at least four elements with identical resistance, while at least one element can be variable and depend on the temperature and be a part of the waveguide, with the other three elements placed at the side of the waveguide's fourth element (**Fig. 1**). At least four metal pads are used to connect the external electrical circuit, including the electrical power source. The external biased voltage is applied between opposite pads, for example**,** of the circuit, while the signal voltage is detected between the other two pads [**29**, **30**].

The monitored element can consist of an absorptive element that is part of the waveguide or the external element that is either part of the waveguide or serves as an additional component that can only slightly contribute to the propagating mode. The external element may be in homogenous form or structured and can be deposited above the waveguide or below, or on any side of the waveguide. Both the absorptive element and the external element are intended to enhance the absorption of light in the waveguide.

The responsivity of the power monitors is, at first, evaluated in the electrical domain by first measuring the signal voltage of the Wheatstone bridge in the absence of the light coupled to the monitored part of the waveguide and then measuring the signal voltage for different amounts of power coupled to it (the monitored part of the waveguide). As it has been previously shown [**29**, **30**], the signal voltage depends on both the input optical power coupled to the waveguide and the applied bias voltage.

**Theoretical background of the Wheatstone bridge arrangement**

The operation principle of the proposed circuit relies on the measurement of the voltage drop through the Wheatstone bridge, which is a direct indicator of the power provided by the optical signal propagating in a given element of the photonic circuit. This voltage is then used to self-calibrate the circuit by using a feedback loop through which voltage is applied to the preceding actuator to stabilize the circuit at the desired value defined by the target voltage.

The optical power absorbed by an absorptive material under a propagation of light in a waveguide is a direct indicator of an amount of light propagating in a waveguide. According to Fourier's Law for heat conduction, the temperature change *ΔT* in a material depends on a power dissipation *P* through a resistive element with absolute thermal resistance $R_{th}$. In terms of a propagating mode, the power dissipated by a material depends on the attenuation coefficient of the propagating mode and the length of the active region, while a dynamic increase of a material temperature *ΔT(t)* due to the absorption of the mode power at time *t* can be expressed as

$$\Delta T(t) = R_{th} P_{in} [1 - \exp(-\alpha_{pr} L)] \cdot \left[1 - exp\left(-\frac{t}{\tau}\right)\right] \qquad (1)$$

where $R_{th}$ is the thermal resistance of the ridge ($R_{th}=L/(\kappa \cdot A)$), *κ* is the thermal conductivity of the waveguide material, *L* is the active length of the waveguide, *A* is the cross-section area perpendicular to the path of heat flow, $P_{in}$ is the power coupled in the mode, $a_{pr}$ is the mode attenuation coefficient (in units 1/μm), and $τ$ is the thermal time constant needed to load thermally a ridge. For comparison, the propagation loss of the DL-SPP plasmonic waveguide was experimentally evaluated at $a_{pr}≈2·10^{-2}$ 1/μm [**29, 30**], while for a photonic waveguide with the a-Si core, it was measured at $a_{pr}≈2.16·10^{-3}$ 1/μm [**31**], which is an order of magnitude lower compared to its plasmonic counterpart.

The temperature rise in a material causes an increase in the material resistivity that can be evaluated under the stationary conditions as follows:

$$R(P_{in}) = R(P_{in} = 0)[1 + \alpha_{th}\Delta T(t)] \tag{2}$$

where $α_{th}$ is the thermal resistance coefficient (temperature coefficient of resistance, TCR) of the material (metal or semiconductor), which, for example, for gold is $α_{th}=3.715·10^{-3}$ 1/K, while for an amorphous silicon (a-Si) it is $α_{th}=2.5·10^{-2}$ 1/K. On the contrary, a low thermal conductivity of a-Si taken at $κ=2.2$ W/(m·K) ensures a good confinement of generated heat in a localization region. A simple way to increase a resistance is selecting a material with a high thermal resistance coefficient $α_{th}$.

Another way to improve the device's performance is to enhance the absorption efficiency of the material that constitutes the waveguide. Transparent conductive oxides (TCOs) in the epsilon-near-zero (ENZ) point seem to be good candidates for this task due to the high enhancement of the electric field in the TCO close to the ENZ point, which translates to a much higher electron temperature rise in a material [**33, 34**]. Furthermore, they are characterized by a low thermal conductivity coefficient evaluated at $κ=0.16-1.2$ W/(m·K) for AZO, which depends on deposition conditions [**31, 32**]. In addition, plasmonic arrangements can be implemented to confine the incident radiation and, as a consequence, improve the absorption efficiency [**31**].

In the Wheatstone bridge configuration and in the absence of the electromagnetic radiation in a waveguide ($P_{in}=0$ W), the signal voltage is expressed by

$$V_s(P_{in} = 0W) = \frac{R_x R_1 - R_2 R_3}{(R_x + R_3)(R_1 + R_2)} V_b \tag{3}$$

Thus, in a perfectly balanced bridge, $R_xR_1=R_2R_3$ and, consequently, $V_s=0$ V.

For the balanced Wheatstone bridge, i.e., when $R_1=R_2=R_3=R_x(P_{in}=0)$ with the $R_x$ resistance being a monitored part of the waveguide in the absence of the light coupled to the waveguide, the signal voltage can be expressed as follows

$$V_s(P_{in}) = \frac{1}{2}\frac{\alpha_{th}\Delta T}{2 + \alpha_{th}\Delta T} V_b \tag{4}$$

where $V_b$ is the bias voltage (**Fig. 1 and 2**) and the scattering contribution is neglected, i.e., it is assumed that all propagation losses go into the absorption losses, which is a reasonable assumption for photonics and plasmonic dielectric-loaded surface plasmon polariton waveguides (DLSPPWs). **Eq. 4** suggests that the changes in the signal voltage with respect to the input optical power are linear, at least in a low regime of optical power coupled to the examined part of the waveguide. It has been confirmed experimentally [**29, 30**], which shows the possibility of monitoring the power-induced changes in the resistance. Furthermore, **eq. 4** suggests that in the limit $\Delta T \rightarrow \infty$, i.e., for a high temperature rise in an examined media, the signal voltage $V_s \rightarrow 0.5 V_b$.

Assuming that two separated waveguides are part of two opposite elements of the Wheatstone bridge, the signal voltage can be expressed as

$$\Delta V_s(P_{in}) = \frac{\alpha_{th}(\Delta T_x - \Delta T_1)}{(2 + \alpha_{th}\Delta T_x)(2 + \alpha_{th}\Delta T_1)} V_b \tag{5}$$

In consequence, for a perfectly balanced bridge, i.e., $\Delta T_x = \Delta T_1$, the signal voltage $V_s = 0$. In comparison, for one of the elements, i.e., "resistors", staying at the environment temperature, i.e., $\Delta T_1 \rightarrow 0$, the signal voltage $V_s$ can be expressed by **eq. 4**. On the contrary, when a temperature of one element of the Wheatstone bridge is much higher compared to the second element, i.e., $\Delta T_x \gg \Delta T_1$, the signal voltage $V_s$ approaches half of the bias voltage $V_b$, i.e., $V_s = 0.5 V_b$.

The spectral response of the MZI varies with wavelength and depends on the beam intensity ratio between both propagating beams along the MZI arms and the phase difference between them; thus, the output power from the MZI is given by

$$P_{out} = \frac{1}{2}(P_1 + P_2)\left(1 + \frac{2\sqrt{P_1 P_2}}{P_1 + P_2} cos(\Delta\phi)\right) \tag{6}$$

where $P_1$ and $P_2$ are the powers in both MZI arms and factor

$$V = \frac{2\sqrt{P_1 P_2}}{P_1 + P_2} \tag{7}$$

also known as a visibility, it defines the beam power ratio between both MZI arms, while $\Delta\phi$ corresponds to the phase difference between them that, for equal length of the interferometer arms, is proportional to the difference in the mode effective indices between two propagating modes, $\Delta n$. The phase difference is expressed by

$$\Delta\phi = \frac{2\pi \Delta n L_{arm}}{\lambda} \tag{8}$$

thus, taking into account the thermo-optic switch, it can be written as

$$\Delta\phi = \frac{2\pi L_{arm}}{\lambda} \frac{\partial n}{\partial T} \Delta T_{arm} \tag{9}$$

where ∂n/∂T is the thermo-optic coefficient of the material that constitutes a ridge, while $ΔT_{arm}$ is the temperature rise in the MZI arm under an applied voltage $V_s$ provided to the actuator from the Wheatstone bridge under measurements.

**Closed-loop control system**

A monitoring device proposed here may be directly implemented in the photonic integrated circuit and may be a part of the photonic integrated circuit. No additional photonic components are required in this arrangement, except some additional electrical ones. To perform a monitoring of any optical signals in a network, some conductive elements are needed that do not influence, however, the optical signals in a circuit.

The examples provided here refer to the programmable forward-only matrix circuit [3, 15, 20] but can also be implemented in any other programmable photonic circuits. Forward here refers to the optical signal flow in a circuit, as referencing may concern both the preceding and following elements.

The signal voltage from the Wheatstone bridge is used to control the actuator, e.g., phase shifting element or modulator, to stabilize the characteristic of interest (e.g., temperature, phase, power) of the photonics component.

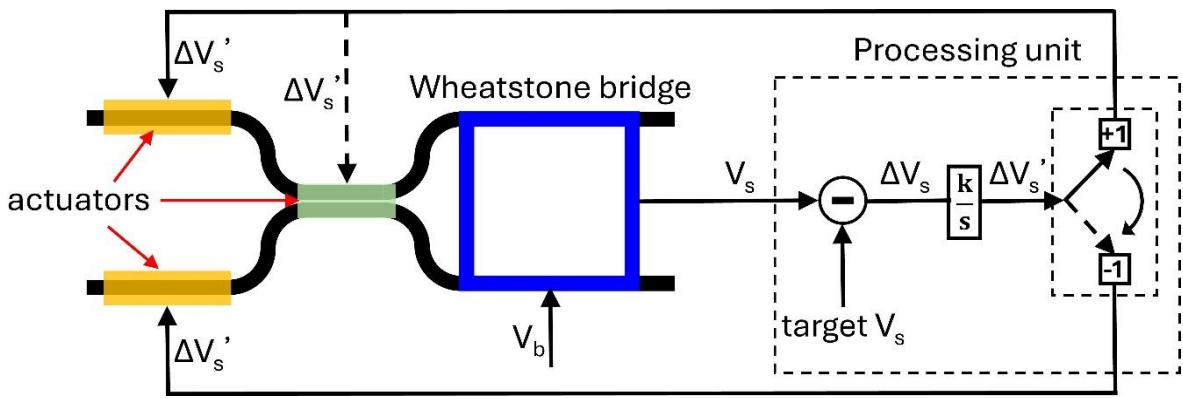

**Figure 3**. Schematic view of the local feedback with the Wheatstone bridge used in the input optical signal generation.

The operation principle of the proposed system was illustrated in **Fig. 3,** where both arms of the MZI serve as the opposite "resistors" of the Wheatstone bridge. In this arrangement, the sign and the strength of the signal voltage are a direct indicator of the beam power ratio between both arms of the MZI. The signal voltage and thus the beam power ratio can be adjusted to a desired value through the feedback loop – the target signal voltage, *target $V_s$*, corresponding to a desired optical power ratio between both arms of the MZI, is subtracted from the measured signal voltage $V_s$, and the resulting difference $ΔV_s$, after an amplification $ΔV_s'$ (if needed), is applied either to one of the phase shifters or the coupler/splitter. Thus, a phase shifter or a coupler/splitter is controlled by the local feedback loop shown in **Fig. 3,** relating the signal voltage $V_s$ to the voltage driving the heater $ΔV_s'$. The voltage from the Wheatstone bridge, $V_s$, is fed to an integral controller that is a part of the processing unit, for example FPGA, that updates the heater voltage $ΔV_s'$ until it is zeroed, thus locking the measured signal voltage $V_s$ to the target voltage, *target $V_s$*. Depending on a sign of the voltage $ΔV_s'$, positive or negative, the resultant voltage can be applied to the phase shifters placed either in a lower or upper arm of the preceding MZI. For example, a positive resultant voltage $ΔV_s'$ can be applied to a phase shifter placed in the upper arms of MZI, while a negative resultant voltage $ΔV_s'$ can be applied to a phase shifter placed in the lower arm of MZI. If a signal voltage $V_s$ from the Wheatstone bridge is equal to a target signal voltage, *target $V_s$*, then the resultant signal voltage $ΔV_s$ is zero, and no voltage is applied to the phase shifters – the MZI remains in its current beam power ratio.

The proposed arrangement can be used to stabilize the working point of PICs against thermal drifts or fabrication inaccuracies by implementing the integrated local feedback loops to drive each photonic device independently at a cost of only a fraction of the power (due to absorption losses in the waveguide under a measurement) and area required by an electronic board. The feedback loop may be designed to both monitor and set the percentage of power on the two outputs of the MZI or any other parts of the photonic circuit.

The feedback control loop is used to drive a signal voltage into the actuator, e.g., a modulator or a phase shifting element, to control a characteristic of the optical signal in the Mach-Zehnder interferometer. Under an applied voltage, the electrode of the actuator generates heat to control the temperature of the waveguide, with the amount of heat generated by the electrode depending on measurements taken by the Wheatstone bridge. The actuator is driven by a signal voltage measured by the Wheatstone bridge that is transferred to it by the processing unit. The processing unit performs simple operations such as subtraction of the measured signal voltage from a target voltage and integration of the resultant voltage to update a heater voltage at the actuator. The proposed configuration allows for the elimination of conventional high-precision digital-to-analog converters (DACs) that are very sensitive to noise, require a large amount of power to operate, are characterized by slow operation speed, and require a large area for implementation.

The monitoring device and actuator are arranged in a closed-loop control system that is controlled through a processing unit, for example, a field-programmable gate array (FPGA), that sets the working point of a photonic device and stabilizes it against external perturbations. Going into details, the readout stage, i.e., the power monitor, extracts the voltage, i.e., a signal voltage, and feeds it to an integral controller in a processing unit, which updates the actuator DC voltage until the output optical signals stabilize in a given condition defined by a processing unit, and thus, effectively locking the system in a desired working point.

Depending on the requirements, the signal voltage from the Wheatstone bridge module may be amplified in a processing unit by the instrumentation amplifier, a high-impedance amplifier arranged in a lock-in detection scheme that allows canceling out the effect of the interconnections and of 1/*f* noise on the measurement accuracy. It should be emphasized that a lock-in readout scheme is preferred for the readout, allowing the minimization of the contribution of the 1/*f* noise of the preamplifier to maximize in this way the sensitivity of the power monitor [**23**]. Simultaneously, the Wheatstone bridge arrangement allows us to minimize the effect of the parasitic series resistance of the bonding wires and pads [**22**, **23**, **28**].

Compared to any other previously proposed devices, the proposed arrangement does not require any observation ports, viewports, that take out of the optical system a small amount of light to perform a detection of light intensity by the opto-electric element, e.g., detector, which increases the complexity of the system [**3**, **6**, **19**, **21**]. In contrast, signal detection is carried out directly on the components of the photonic circuit that are directly involved in signal transmission. Therefore, it does not require any additional photonic components.

1. **Monitoring a power in one arm of the MZI**

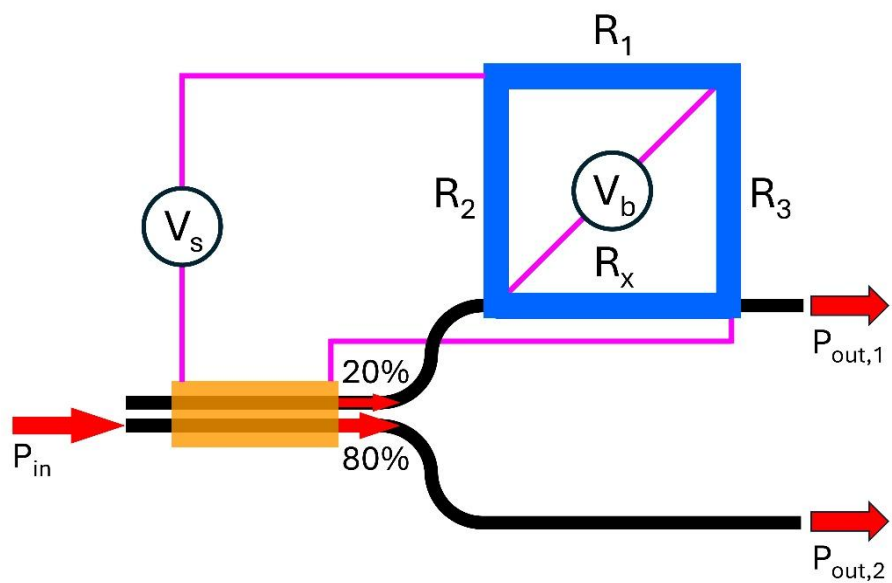

**Figure 4**. Schematic view of the local feedback with the monitoring device in the form of a Wheatstone bridge arranged in one of the arms of the MZI and an actuator placed in a coupler/splitter section.

The proposed monitoring device can be implemented in any part of the photonic system, photonic mesh, and can provide a resultant voltage to any part of the system both preceding and following the monitor. The Wheatstone bridge implemented in one of the arms of the MZI may be exploited to control a signal in the MZI arms and, especially, to maximize the amount of light propagating at the waveguide, as shown in the example in

**Fig. 4**. This is obtained by tuning the coupler/splitter heater voltage by the signal voltage to drive the output signal at the waveguide to the value defined in a processing unit.

Compared to other techniques, this arrangement does not require any prior calibration and is not affected by any other parameters of the system, such as instabilities of the input light power or temperature drifts. The coupler/splitter voltage is automatically set by the Wheatstone bridge signal voltage that is fed with an optical signal propagating in a waveguide. The signal voltage is adjusted to provide a negative feedback loop only when some optical signal propagates in the waveguide, thus avoiding ambiguity on the target locking point.

To serve only as an example, let's assume that in initial conditions the light is coupled to waveguide and then is split in a splitter in a ratio of 80/20, with 80% of the light coming to a direct waveguide and 20% coming to a crossing waveguide (**Fig. 4**). The Wheatstone bridge is placed in the crossing waveguide arm, where it is used to monitor an amount of light in this waveguide through the active part of the waveguide. Under an absorption of light by an active element its temperature rises. By applying the bias voltage between the opposite sides of the Wheatstone bridge, the signal voltage appears that depends directly on the bias voltage and the amount of light coupled to the waveguide. Then, the signal voltage is applied to the electrode placed in a splitter section. Under an applied voltage, the electrode is heated, and heat is transferred to the waveguides in a coupling section. The temperature of waveguides increases, which provides a change in their refractive indices through the thermo-optic mechanism and, in consequence, a change in a coupling ratio between waveguides. As a result, more light couples to the crossing waveguide. The more power in a crossing waveguide means the higher signal voltage drop in the Wheatstone bridge and, in consequence, the more voltage applied to the heating electrode in a coupler/splitter section, which means further increases of light coupled to the crossing arm. The process will continue until the entire light comes to a crossing waveguide. No further increase in a signal voltage can be expected for an established bias voltage, as no additional increase in the power coupled to the crossing arm is possible. The explanation above applies to cases where there is no target voltage set in the processing section.

For a second example, let's assume the splitting ratio of 100/0 in a splitter, i.e., when all light is coming to a direct arm and no light is observed in a crossing arm, the signal voltage from the Wheatstone bridge placed in a crossing arm shows zero voltage drop as no power is absorbed by the active element of the waveguide. Thus, no power is applied to the electrode in a splitter section as a bridge is totally balanced, i.e., all resistors are the same, and no change in the splitting ratio between a direct and crossing waveguide can be expected. It means that the entire light is still coming through a direct arm. As in previous case, it refers to the case where no target voltage is set in the processing unit. Using the processing unit it is possible to redirect light to the crossing waveguide in the desired proportion by setting the target voltage accordingly. In this case, even if the Wheatstone bridge produces zero $V_s$, the voltage supplied to the actuator will be the voltage corresponding to the target voltage. Once the desired proportion of light in the waveguides has been achieved, the system can be locked at the corresponding voltage either by using non-volatile materials or by applying that particular voltage to the actuator.

2. **<u>Monitoring a power in a termination of the MZI</u>**

The proposed Wheatstone bridge can be placed in any part of the photonic system and can be arranged in a closed-loop with any actuators, both preceding and following a monitor.

Using another example, let us assume that the Wheatstone bridge is placed in a termination waveguide of the MZI (that can serve simultaneously as one of the arms of another MZI) while the actuator(s), arranged in a closed-loop control system, is placed in one of the preceding arms (or both) of the MZI (**Fig. 5**). This arrangement can be used to stabilize the system at the desired operating point even in the presence of external perturbations. The control strategy relies on a measurement of the signal voltage $V_s$ of the Wheatstone bridge and then applying it to the actuator(s), thus modulating the working point of the MZI. The actuator DC voltage is updated until the signal from the termination waveguide of the MZI is zeroed, effectively locking the MZI at the point of two beams propagating out-of-phase in the opposite arms of the MZI. Again, it refers to a case where no target voltage is set in a processing unit.

The occurrence of this phenomenon is contingent on the presence of sufficient signal voltage, capable of inducing a $\pi$ phase shift between the two arms of the MZI. The signal voltage can be adjusted by controlling either the optical power supplied to the MZI or the bias voltage $V_b$ applied to opposite sides of the Wheatstone bridge. In the instance of a low bias voltage $V_b$, the signal voltage $V_s$ drops, and the MZI is entrapped in a state of minimal transmission corresponding to a specific phase shift between the two arms of the MZI. Therefore, the phase shift between the two arms of the MZI can be controlled and locked using the bias voltage $V_b$.

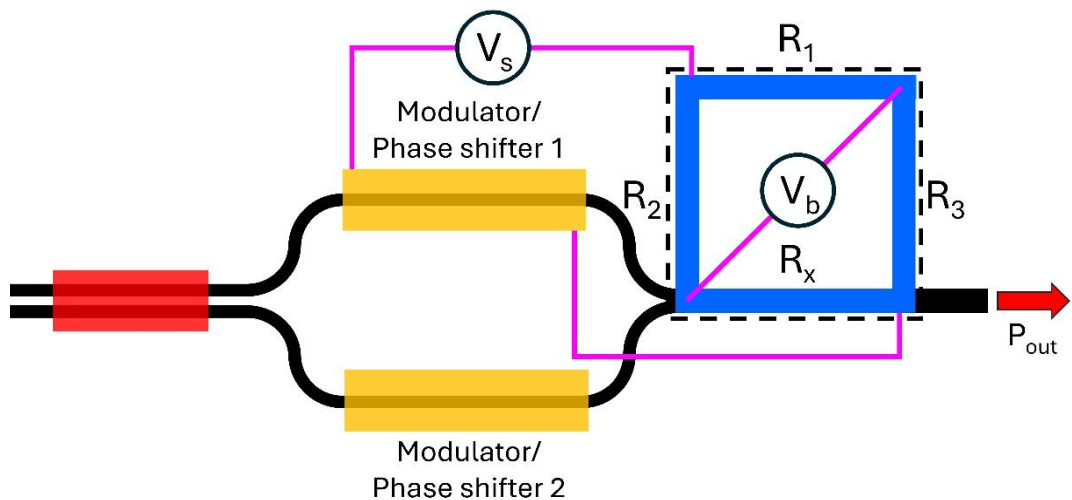

**Figure 5**. Backward referencing device with the Wheatstone bridge implemented in a termination of the Mach-Zehnder Interferometer (MZI).

The signal output from the interferometer can be adjusted using two actuators located in both arms of the MZI to maximize or minimize the signal output from the waveguide. For an equal beam intensity ratio at the output of both arms of the interferometer and for both signals propagating in phase, the constructive interference of the optical signals at the termination of the interferometer occurs, and power at the output waveguide is maximized. This raises the temperature of the active component of the waveguide and consequently alters its resistance, which is thus distinct from the other three resistances comprising the Wheatstone bridge. Consequently, the Wheatstone bridge moves out of equilibrium, resulting in an increase in the signal voltage. The resulting signal voltage is then applied to one of the preceding actuators, thereby modulating the optical signal at the output of the selected interferometer arm.

The signal voltage exhibits a linear dependence on both the optical power absorbed by the active element of the waveguide, bridge, and the bias voltage. Therefore, by increasing the bias voltage, it is possible to increase the signal voltage even with low optical power propagating through the waveguide. When the signal voltage applied to the preceding actuator is sufficient to introduce a $\pi$ phase shift between the two arms of the interferometer, destructive interference occurs and the signal at the output of MZI drops to its minimum value. Consequently, with no power at the termination of the interferometer, all the bridge resistors exhibit the same resistance, resulting in the bridge being balanced and the signal voltage dropping to zero. This leads to a lack of voltage supplied to the preceding actuator, resulting in a return of the system to its initial state. In the absence of voltage supplied to the actuator, the signals propagating in both arms of the interferometer are in phase, resulting in a maximum signal at the output of the MZI. As a result, the temperature of the absorptive element of the output waveguide rises again, bringing the bridge back out of equilibrium.

In the absence of the electronic controller in a processing unit module, that may control the electrical signal applied to the actuator(s) and, finally, lock it on a preferred value, the system oscillates between the two states with an oscillation frequency determined by the thermal time constants of both the phase shifting element/modulator and the absorptive element of the output waveguide, which is a part of the Wheatstone bridge. In consequence, the output signal from a system changes from zero to a maximum value defined by the amount of optical power provided to the system.

An electronic controller arranged in a feedback loop can control the voltage supplied to an actuator by adjusting the signal voltage from the Wheatstone bridge to the target voltage set on the controller. Consequently, a constant voltage is applied to the actuator, ensuring stable modulation of light in the waveguide and enabling stable operation conditions of the system.

### 3. <u>Monitoring a power into two arms of the MZI</u>

As previously demonstrated in **Fig. 3**, the Wheatstone bridge can be configured so that the two arms of the interferometer act as the two electrodes of the bridge, with the remaining two electrodes providing the connection between them (**Fig. 6**). The arms of the interferometer serve simultaneously as the couplers/splitters from other interferometers that combine the optical signals coming from different inputs. The upper arm of the interferometer combines the optical signals from upper MZIs, while the lower arm combines the optical signals from lower MZIs. The preceding actuator, i.e., a phase-shifting element or modulator, is arranged in a coupler/splitter section that provides signals to the opposite arms of the interferometer. As the perpendicular electrodes are not affected by a propagating light, the Wheatstone bridge can only be affected by electrodes that are part of the interferometer.

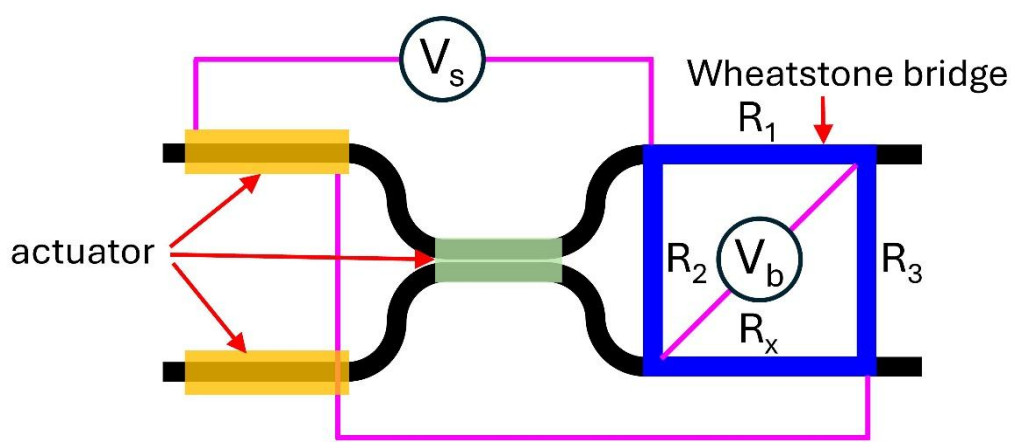

**Figure 6**. Schematic view of the local feedback with the monitoring module in the form of the Wheatstone bridge, where two arms of the MZI serve simultaneously as two elements of the Wheatstone bridge, and with the actuator placed in a preceding coupler/splitter section that may serve simultaneously as one of the arms of the preceding MZI.

For the same beam intensity ratio between both arms of the interferometer, the signal voltage $V_s$ is zeroed as indicated by *eq. 5*. For a higher power in the lower arm compared to upper arm, more heat is dissipated in the Wheatstone bridge "resistor" $R_x$ compared to "resistor" $R_1$ and a signal voltage $V_s$ is positive. The signal voltage reaches maximum positive value when an optical power ratio between both arms is 100/0, i.e., total power is directed to lower arm while there is no power in the upper arm. When the signal voltage shows a negative value, it means that more power is coming through the upper arm compared to lower arm. The maximum negative value of the signal voltage is achieved when total power is coming through the upper arm while there is no power in the lower arm, i.e., the interferometer optical power ratio is 0/100.

**Material platform for a realization of the control-loop**

One of the possible realizations of the Wheatstone bridge in the photonic integrated circuits can be based on the plasmonic platform, where a metal stripe supporting a plasmonic mode can serve simultaneously for a monitoring of power in a plasmonic waveguide due to the mode absorption [**29, 30**]. In this case, the monitoring device is based on a metal supporting a plasmonic mode.

The measurements performed for a monitoring of an optical power propagating in a plasmonic mode revealed very high responsivities reaching up to 6.4 µV/µW for a bias voltage of only 245 mV. It was an impressive result even assuming a very small thermal coefficient of resistance for gold assumed at $α_{th}$=3.715·10$^{-3}$ 1/K (~ 0.37 %/K) [**29, 30**]. In comparison, the responsivity for a previously reported in-line plasmonic photodetector was measured at 7.5 µV/mW, which is over three orders of magnitude lower compared to the Wheatstone bridge arrangement [**23**].

Apart from it, the same metal stripe can be used as an electrode, allowing for heating the plasmonic device; thus, it allows for a realization of active plasmonic components such as, for example, plasmonic-based phase shifters [**35**]. In consequence, depending on the requirements, the same arrangement can be used for both a monitoring of power in the plasmonic waveguide and as a heater in phase shifting elements. Unfortunately, the main problem in plasmonic technology stems from high losses related to metal-induced attenuation. This impact can be minimized by the integration of short plasmonic active components with longer passive photonic waveguides [**35**].

An alternative material platform for a realization of on-chip Wheatstone bridges can be based on the a-Si platform. a-Si is a common material for bolometers exhibiting a high temperature coefficient of resistance (TCR) of 2.5 %/K and simultaneously a low thermal conductivity of 2.2 W/(m·K). However, it suffers from a high resistivity; thus, a change in resistance is difficult to measure. To mitigate this problem, a heavily doped p-type a-Si layer is usually placed on top of the waveguide to reduce the overall resistance. In consequence, a waveguide with a 400 nm thick intrinsic a-Si layer and a 100 nm thick p-type a-Si layer on the top was proposed for the bolometric applications [**31**]. For the Wheatstone bridge configuration, this material stack can be a part of the monitored element of the waveguide, while the rest of the waveguides can be based on a standard Si platform. The connections between monitored parts of the waveguides can be realized based on the metallic stripes.

In terms of the actuators, the thermo-optic phase shifters used within each MZI can rely on the resistive Titanium Nitride (TiN) heaters placed several microns directly above the waveguide. To provide an efficient heat transfer to the below waveguide, each heater should be at least 2 µm wide and 120 µm long. Furthermore, to provide a level of mitigation towards thermal crosstalk, the thermal isolation trenches can be etched through the buried oxide and into the silicon substrate on either side of each heater [**36**, **37**].

Another possible realization of the Wheatstone bridges on a chip can be based on the transparent conductive oxide (TCO) materials that can be deposited on top of the waveguides [**33**]. A thin layer of TCO material, for example, ITO, AZO, GZO, etc., can be deposited on top of the waveguide with a separation between the waveguide and TCO provided by a thin oxide layer. The TCO material properties can be chosen or actively modified to be close to the epsilon-near-zero (ENZ) point that ensures enhanced absorption in the TCO. Furthermore, due to a low carrier concentration, a much higher electron temperature rise in a material can be achieved [**33**]. Apart from it, they are characterized by a low thermal conductivity coefficient evaluated at $\kappa$=0.16-1.2 W/(m·K) for AZO, which ensures a good confinement of a generated heat in a localization region [**38**].

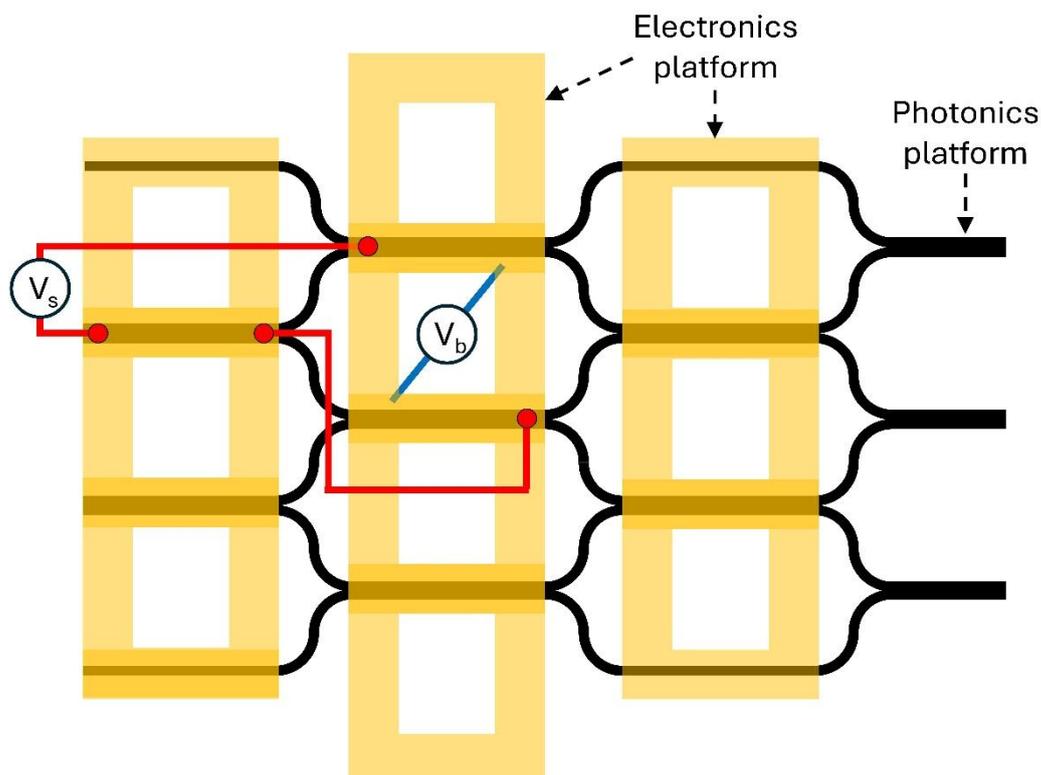

**Figure 7**. A schematic representation of the proposed system incorporating an electrical circuit on top of the photonic feed-forward mesh. The electrical circuit can be utilized to monitor the optical power in any section of the mesh, and it can also be employed as an electrode to modulate light in any particular section of the mesh. Consequently, the establishment of the control loop is feasible at any point on the mesh structure.

Being optically transparent (i.e., low-loss) and electrically conductive, TCOs can serve simultaneously as a heater that can be directly integrated on a photonic device [**38**]. The TCO layer can be electrically driven via two gold/titanium (Au/Ti) electrodes formed on its two ends. The electrical conductivity of the TCO layer is an important design parameter that directly affects the efficiency of the Joule heating and provides small additional optical loss to a device. To electrically isolate material from TCO film, a thin layer of oxide spacer can be deposited between the material and the TCO heater [**38**].

Therefore, depending on the requirements, the same TCO arrangement can be used for both monitoring an optical power in a waveguide and for activating an actuator through the TCO heater to provide a heat transfer to a phase shifting element or modulator.

Consequently, each component of the mesh can be concurrently monitored and activated through the same configuration. As illustrated in **Fig. 7**, for a feed-forward mesh architecture, the photonic platform constitutes the first layer, while the second layer represents the electrical circuit composed of TCO. However, the proposed arrangement can be as well implemented for the recirculating meshes, especially to the square mesh and a recently proposed shifted rectangular mesh, so called "bricks" mesh topology [**39**].

Thin layers of TCO stripes are placed on each waveguide, while the connections between them are made vertically by other TCO strips or, alternatively, by electrodes made of alternative materials. The configuration of the system establishes a Wheatstone bridge that can be utilized to monitor the signal propagating in each waveguide. In this scenario, the bias voltage is applied to the opposite pads of the bridge, while the signal voltage is detected using the other two pads. The TCO layers deposited on the waveguide can function as heating electrodes when a voltage is applied to both ends of the electrode [**38**]. Consequently, each TCO electrode deposited on the waveguide is capable of performing dual functions. Firstly, the device has the capacity to function as a component of a Wheatstone bridge, thereby enabling the monitoring of signals propagating within a waveguide. Secondly, it can serve as a heating electrode, which is employed to modify the signal propagating in the waveguide.

It is worth to mention here that a recent geopolitical tensions between the major powers, the war between Russia and Ukraine, as well as economic and social impact of the COVID-19 pandemic revealed vulnerabilities in some global economies due to interrupted supply chains. Thus, there is a pressing need among many world economies to build more resilience and strategic autonomy in supply chains that are critical for stable economy and security. Some of the most popular TCO materials, for example ZnO compounds, seem to fit very well into these requirements as they are made of abundant natural elements often occurring in nature. For this reason, they are cheap what ensures low-cost fabrications.

The application of the signal voltage from the Wheatstone bridge, which is employed to monitor power propagating in a waveguide, to a preceding actuator results in the creation of a control loop. This control loop is capable of stabilizing a signal at a predetermined level, as specified by the processing unit. Consequently, each location on the photonic mesh can be subjected to monitoring and automated stabilization at a designated level, with the utilization of only a negligible optical signal.

**Conclusion**

Here we proposed a new type of closed-loop stabilization and calibration system that can operate in real-time and at each part of the photonic mesh structure, including both a feed-forward and recirculating mesh architecture. The system is based on a Wheatstone bridge, which is capable of monitoring the optical signal propagating in any part of the mesh. Due to the fact that all bridge components are located on the same chip and in proximity to each other, the system shows little sensitivity to any changes in the temperature of the photonic components located on the chip. Therefore, the only factor affecting the actual temperature change of the tested element is the optical power propagating in the tested section of the waveguide.

Furthermore, in contradistinction to the majority of monitors and photodetectors, which provide electrical current as an outcome of measuring an optical signal, the proposed arrangement provides voltage, thereby eradicating the necessity to convert current to voltage at a later stage. This serves to firstly streamline the entire circuit and secondly contribute to a substantial diminution in the noise level in the circuit.

This approach is expected to unlock new degrees of freedom in programmable photonic circuits, offering almost real-time stabilization and calibration of the system. Furthermore, it paves the way for advanced applications in the fields of topological photonics, quantum information processing, neuromorphic and high-speed optical computing.

**Acknowledgement**

The author acknowledges the constant support of Warsaw University of Technology, Poland, for the completion of this work. Furthermore, he is very thankful to Prof. D. G. Misiek for his support and very valuable suggestions.